\documentclass[]{interact}

\usepackage{epstopdf}
\usepackage[caption=false]{subfig}

\usepackage[longnamesfirst,sort]{natbib}
\bibpunct[, ]{(}{)}{;}{a}{,}{,}
\renewcommand\bibfont{\fontsize{10}{12}\selectfont}

\theoremstyle{plain}

\theoremstyle{definition}

\theoremstyle{remark}

\begin{document}


\title{A Red Teaming Framework for Securing AI in Maritime Autonomous Systems}

\author{
\name{Mathew J. Walter$^{1}$\thanks{CONTACT Mathew J. Walter Email: mathew.walter@plymouth.ac.uk}, Aaron Barrett$^{1}$ and Kimberly Tam$^{1,2}$}
\affil{$^{1}$University of Plymouth, Plymouth, UK, PL4 8AA}
\affil{$^{2}$The Alan Turing Institute, British Library, London, UK, NW1 2DB}
}

\maketitle

\begin{abstract}
Artificial intelligence (AI) is being ubiquitously adopted to automate processes in science and industry. However, due to its often intricate and opaque nature, AI has been shown to possess inherent vulnerabilities which can be maliciously exploited with adversarial AI, potentially putting AI users and developers at both cyber and physical risk. In addition, there is insufficient comprehension of the real-world effects of adversarial AI and an inadequacy of AI security examinations; therefore, the growing threat landscape is unknown for many AI solutions. To mitigate this issue, we propose one of the first red team frameworks for evaluating the AI security of maritime autonomous systems. The framework provides operators with a proactive (secure by design) and reactive (post-deployment evaluation) response to securing AI technology today and in the future. This framework is a multi-part checklist, which can be tailored to different systems and requirements. We demonstrate this framework to be highly effective for a red team to use to uncover numerous vulnerabilities within a real-world maritime autonomous systems AI, ranging from poisoning to adversarial patch attacks. The lessons learned from systematic AI red teaming can help prevent MAS-related catastrophic events in a world with increasing uptake and reliance on mission-critical AI.
\end{abstract}


\section{Introduction}
Artificial intelligence (AI) is becoming increasingly prevalent in the automation of processes abound in science and industry. Despite the uptake, academics are highlighting the threats of AI, from the misuse of AI to the inherent vulnerabilities associated with AI systems \citep{mirsky2022threat, kong2021survey}. The academic literature suggests a number of attacks are possible \citep{szegedy2013intriguing, goodfellow2014explaining, biggio2014security, biggio2013security, mirsky2022threat, kong2021survey, qiu2019review,  walter2023adversarial}, and the existence of security vulnerabilities in AI has given rise to the concept of adversarial AI (AAI) as a descriptor for its misuse. The realms of AAI and eXplainable AI (XAI) have highlighted some of the inherent hazards and safety concerns associated with AI systems, which, in certain instances, may even pose a risk of fatalities \citep{caruana2015intelligible}. Microsoft's Tay chatbot was one of the earliest examples of an attack manipulating the AI's continuous learning process \citep{wolf2017we}. Later, a survey of AI operators suggests evidence that other AAI attacks may have already been deployed in the real world \citep{grosse2023machine}. At a government level, key cyber security leaders have raised concerns about AAI attacks and the lack of secure by design AI systems \citep{bbcMustHave}. In November 2023, the US government called for AI red teaming for mission-critical AI systems \citep{TheWhiteHouse}. With an exponential-like uptake in AI, more reliance on AI for critical decision making and more effective AAI methods, the once considered theoretical threat of the future is fast becoming the present-day threat.

As noted in the work of \cite{walter2023adversarial}, attack and defense methods evaluated in low-entropy laboratory environments often have different effects when utilized in more complex and dynamic real-world environments, where AI is ultimately expected to operate. In computer vision, many attacks also focus on the area of image classification rather than object detection and are applied to low-resolution pixelated images, yielding methods that are rather limited in the applications of marine AI. Therefore, the understanding of the effects of AI-based attacks on mission-critical systems is fairly unknown 
and there is a lack of frameworks 
AI developers/users can use to evaluate AI cyber security. In this work, we propose one of the first red team (adversary simulation) AI frameworks (RED-AI) to evaluate real-world AI systems to try to comprehend the realistic threats to AI systems. 

This RED-AI framework methodology for AI evaluation (Appendix~\ref{append:checklist}) can be tailored for a range of AI system evaluations, however we 
specifically apply the RED-AI framework to a maritime autonomous system (MAS) as a real-world test case in this paper. MAS developers are increasingly embracing AI solutions, and attacks on these systems can compromise the objectives of the operation, making them a practical real-world platform to examine this methodology. 
As of the time of writing, the authors are not aware of an AI security framework for the evaluation of MAS existing in the public domain, making it a useful test-case.
This work aims to apply and examine the proposed RED-AI framework to secure real-world AI; therefore, we do not need to rigorously examine every attack method from the literature, but demonstrate some attacks to highlight the effects of different vulnerabilities on the AI system.

This paper offers the following novel contributions:
\begin{enumerate}
    \item Propose one of the first frameworks (RED-AI) aimed at red teaming/evaluating real-world AI systems. 
    \item Apply the framework to find vulnerabilities in real-world maritime autonomous AI systems and show the effectiveness of the framework. 
    
\end{enumerate}

The paper is structured as follows: We first provide a critical review of the existing literature to motivate the work and understand the current state of the art in Section~\ref{sec:Background}. In Section~\ref{sec:Method}, we then consider the methodology for developing the RED-AI evaluation framework and provide a high-level overview of the framework. Then, in Section~\ref{sec:TestCase}, we evaluate RED-AI against a real-world MAS as a test case. Finally, we provide a discussion in Section~\ref{sec:Discussion}, and later conclude the work in Section~\ref{sec:Conclusion}. The RED-AI framework checklist to secure AI can be found in Appendix~\ref{append:checklist}.

\section{Existing Background}
\label{sec:Background}

AI can be misused for a range of malicious purposes such as to support social engineering and phishing attacks \citep{seymour2016weaponizing, yamin2021weaponized, mirsky2021creation}, biometric spoofing \citep{dang2020detection}, malware development \citep{kolosnjaji2018adversarial,maiorca2019towards, al2018adversarial, li2021arms}, automated hacking/penetration testing \citep{ghanem2019reinforcement, chen2023gail, pozdniakov2020smart}, attacks on cloud services \citep{juuti2019prada}, proposal of novel chemical and biological weapons \citep{urbina2022dual} amongst other nefarious applications. However, this work focuses on the inherent vulnerabilities of complex neural networks \citep{szegedy2013intriguing, goodfellow2014explaining}. These types of threats have been shown to exploit neural networks to achieve model evasion \citep{szegedy2013intriguing, goodfellow2014explaining}, poisoning \citep{ramirez2022poisoning}, stealing and extraction attacks \citep{song2022survey, zhao2021exploiting}. We discuss these vulnerabilities in more detail in Section~\ref{sec:TestCase}.

In previous work \cite{walter2023adversarial}, the authors note that many of these existing attacks and defenses have been performed in limited low-entropy laboratory environments, and to have the greatest impact and truly understand the effects, one must evaluate these methods in the AI's true, usually more complex, nosier and dynamic, real-world environments. This is a primary reason why the real-world threat landscape for AAI attacks is poorly understood. Thus, in this work, we propose an evaluation towards understanding and securing against the real-world threat landscape by considering attacks in real-world environments (MAS at sea). 

Google has attempted to red team AI systems in the work of \cite{blogGooglesTeam}, where they briefly describe possible attack types that could be used and encourage the use of evaluating AI. The works of \cite{jing2021artificial, villegas2023toward} also provided a high-level overview of what an assessment may include but do not attempt to employ this assessment on AI technology. One potential issue with this approach is that the theoretical to practical boundaries are not well defined. The work of \cite{yoo2023formulating} considers the requirements for engineering secure AI MAS.

Most of the relevant existing literature considers the importance of secure by design AI and frameworks against a specific type of attack, i.e., privacy. For example, the framework of \cite{caroline2021securing} attempts to secure by design AI technologies, which, whilst we strongly encourage, does not necessarily support the security of already existing AI technologies in which it may not be practical to recreate with the framework. Further, whilst securing the development of AI is an important step, it is equally important to evaluate the AI with the most up to date threat intelligence, in a reactive manner, on a regular basis. As mentioned, there are also frameworks to protect against a specific type of attack, such as preserving data privacy \citep{khowaja2022toward, fritchman2018privacy}. There are also some existing academic publications that consider an evaluation carried out on specific algorithms such as support vector machines (SVMs) \citep{biggio2014security}, pattern classifiers \citep{biggio2013security} and large-language models \citep{ge2023mart}. 

To the best of the author's knowledge, there is no existing framework to evaluate the security of MAS against adversarial AI. However, more general cyber-risk frameworks have been published that considered the future cyber-risks of MAS by examining three MAS prototypes \citep{tam2018cyber}.

\section{Methodology}
\label{sec:Method}

\subsection{Evaluation Framework}
\begin{figure}[t!]
    \centering
    \includegraphics[width=0.65\textwidth, clip=true, trim=6.00cm 15.20cm 6.0cm 0.80cm]{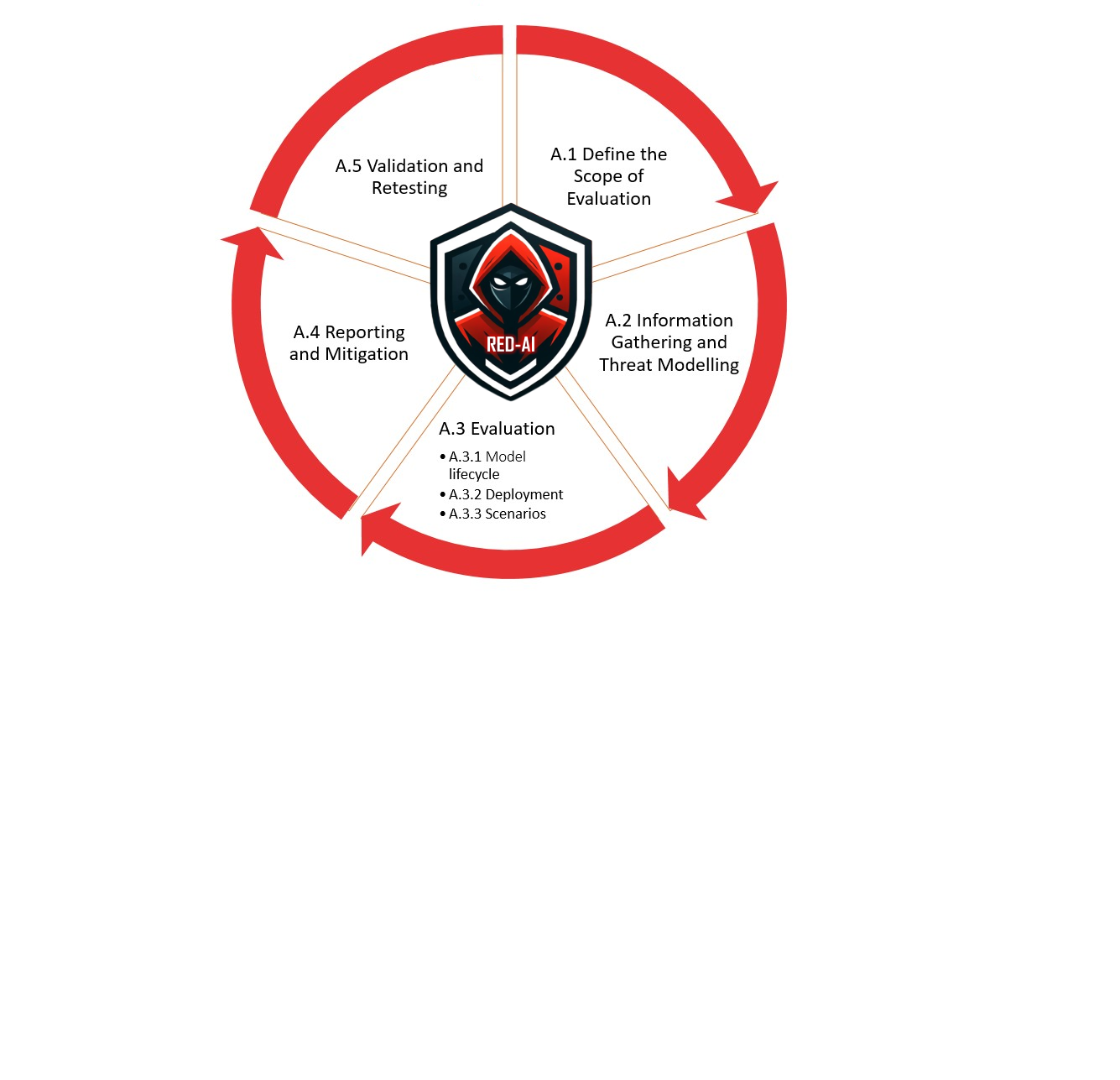}
    \caption{The five step RED-AI framework to red team AI (A.1 $\rightarrow$ A.5).}
    \label{fig:REDAIDiagram}
\end{figure}

In this section, we propose a methodology for evaluating AI. The authors have consulted AI owners and cyber security leaders on the methodology to provide additional insight and recommendations in its development. RED-AI was designed to be as comprehensive as possible to cover a wide range of security aspects, yet general enough to be multi-domain. RED-AI was developed and updated during real-world testing, and we showcase highlights from a real-world example in Section~\ref{sec:TestCase}. An overview of the RED-AI framework is shown in Figure~\ref{fig:REDAIDiagram}.

RED-AI is a general framework and can be adapted to the specific evaluation requirements and types of AI. For example, suppose one is only evaluating a computer vision situation awareness system for a remote autonomous vessel (partially utilizing AI). In that case, this may not require the same depth as evaluating a fully autonomous vessel with many integrated technologies. We note, in some areas, the structure of the evaluation forms a similar structure to a penetration test, which is also used for conventional cyber security evaluation, and this framework could be used alongside conventional penetration testing frameworks for relevant infrastructure where we see systems using a hybrid of AI and non-AI technologies. However, a full conventional security evaluation is outside the scope of this AI-focused evaluation. We start by describing the five key stages (Section \ref{subsec:scope}-\ref{subsec:report}) and later provide a checklist (see Appendix~\ref{append:checklist} for the concise evaluation RED-AI framework checklist).

\subsubsection{Define the Scope of Evaluation}\label{subsec:scope}
Like a penetration test, it is important first to define the boundaries of the evaluation, consider any risks during the test and ensure a mutual understanding of the evaluation with the AI owner. One should do this by defining the evaluation's scope, discussing the evaluation's objectives, looking to minimize any possible disruption to the AI owner, discussing limitations and assumptions, as well as access (including physical) to the systems. 
Furthermore, the rules of engagement should be defined, and the disclosure process of vulnerabilities should be properly settled with the AI owner before enacting the evaluation. 

\subsubsection{Information Gathering and Threat Modeling}
We next need to gather relevant intelligence to develop a threat model for the AI system, which will support one in optimizing the evaluation and highlighting promising areas of compromise; this is to develop an intelligence-directed evaluation rather than brute forcing many unnecessary and costly methods.
We consider both external publicly available data through open source intelligence (OSINT) and internal data relating to the owner's AI. This phase relates to the collection of relevant information for threat modelling and acquiring information that can be utilized in the evaluation and scenario phase (for example, creating proxy models). 

\subsubsection{Evaluation}
In the evaluation phase, we utilize our understanding of the threat model to consider how attacks may affect the security of the AI. This phase could be considered similar to the exploitation stage of a penetration test.
We will evaluate the security of the model throughout its lifecycle from development to deployment and decommissioning. 
We could also consider AI-related cyber security attacks, such as attempting to bypass security measures implemented within the AI system (with prompt injection), the effects of attacks on different sensors and AI systems through to the four different categories of adversarial AI attacks: extraction, evasion, inversion and poisoning. We also consider realistic delivery methods for relevant attacks and develop unconventional malware to execute these actions.

One should consider closed-box and open-box attacks during the evaluation phase, which will depend on the access granted by the AI owner set out in an earlier phase (RED-AI \ref{App:DefineScope}). Open-box (also known as white-box) refers to an attacker having complete access to a model's parameters, and closed-box (also known as black-box) refers to an attacker having no access to the model's internals. We note that some attacks, such as extraction attacks, can support and improve other attacks, such as an inversion attack or allow the use of open-box attacks. 
One may examine and consider the effects on the AI model of denial-of-service (DoS) of sensors and communications.

\subsubsection{Scenarios}
In addition to the evaluation stage, we consider the possibility and likelihood of the previously evaluated attacks and threats. This will allow the user to prioritize and address the threats in a systematic way. By considering the ease of attack, the gain to the attacker and the effects of the attack, the user can categorize the relevant threats to their system. This will need to be tailored for the specific type of AI and its application, as the effects of different attacks vary significantly between applications. We could also consider misuse of the technology, particularly in evaluating emerging AI technologies. 

\subsubsection{Reporting and Mitigation}\label{subsec:report}
When the threats have been identified and tested to work in the AI's real-world environment. We then suggest mitigating defenses and provide a report (and a meeting) of the actions conducted during the evaluation to communicate the findings of this evaluation. The report should include disclosure of potential vulnerabilities, recommended mitigation and possible effects of attacks and could include threat scenarios. We will consider the UK's National Cyber Security Center (NCSC) principles for the security of
machine learning \citep{anonymous_2022} to provide appropriate mitigations and practices. One should also highlight any areas they were not able to evaluate thoroughly and any assumptions made during this test. 

\subsubsection{Validation and Retesting}
After the report, the AI owner should be aware of the risks and vulnerabilities of the AI system and lifecycle; they should also be aware of the mitigations available and may want to action these mitigations and validate the changes with a retest.
Furthermore, as the threat landscape evolves, it becomes appropriate to retest and revalidate the AI, defences and mitigations periodically against new and existing adversarial AI threats.

\subsection{Maritime Autonomous Systems}
In this work, we assess a maritime autonomous system (MAS) against the threat of adversarial AI. We demonstrate the utility of the proposed AI evaluation framework, RED-AI, by applying the assessment to real-world AI technology. One vital sector growing in the utilization of AI is the maritime sector, which accounts for 
around 90\% of international trade \citep{kaluza2010complex}. Evaluating a real-world MAS platform allows one to identify the real-world effects of adversarial attacks because the model operates in complex and dynamic environments (for example, the sea and busy ports). The dynamic operating conditions cause a range of challenges from variations in light, reflections, other interfering vessels and complex environmental conditions, which have large effects on the efficacy of attacks; it is, therefore, vital to evaluate these technologies in the real-world environment to gauge the true impact. 

\begin{table}[b!]
\tbl{IMO definitions for Autonomy Classification Levels.}
{\begin{tabular}{ll} \toprule
 \textbf{IMO Level} & \textbf{Description}  \\ \midrule
 1 & Considers systems which contain autonomy to support the vessel's crew.  \\
 2 & Systems are crewed but can be controlled remotely.  \\
 3 & Systems are uncrewed and controlled remotely.  \\
 4 & Systems are uncrewed and fully autonomous. \\ \bottomrule
\end{tabular}}
\label{tab:IMO}
\end{table}

Maritime autonomous systems have been categorized into four classifications of autonomy by the International Maritime Organization (IMO), from IMO level one, which considers systems that contain autonomy to support the vessel's crew, to IMO level 4 systems, which are uncrewed and fully autonomous vessels. In this work, we consider a vessel which sits between level 3 and level 4, as it can be controlled from a remote operations centre (ROC) and by AI. See Table~\ref{tab:IMO} to observe IMO levels in a table format. Most currently deployed MAS sit in the IMO level 3 category, which allows the crew to operate craft remotely, providing benefits over conventional vessels such as crew physical safety, better working conditions (less time at sea), reduced vessel development, crew costs and allows for more efficient and economically friendly transport of cargo \citep{askari2022towards, porathe2014situation, morris2017worlds, ziajka2021costs, munim2019autonomous, kretschmann2017analyzing, felski2020ocean, tsvetkova2022creating}. However, IMO level 4 vessels can further enhance these benefits and mitigate some cybersecurity challenges resulting from the hacking, spoofing, and jamming of remote communications \citep{fan2020framework}.

Furthermore, AI-driven systems may have significant technical benefits over IMO level 3 systems, such as reacting without emotion, making more statistical-based decisions, reacting much faster, and providing complete autonomy. These benefits in a military domain provide a significant strategic advantage over anyone using conventional systems, a powerful force multiplier, and governments need to maintain a competitive military edge to deter aggression. Thus, we are seeing a drive for higher levels of autonomy for vessels in both civilian and military domains. Development of these systems is ongoing with some real-world examples of MAS, including the Mayflower autonomous ship, which crossed the Atlantic Ocean autonomously in 2022 and intends to be a fully IMO level 4 automated vessel from utilizing AI \citep{anderson2019bon}. The Norwegian Yara Birkeland project has also demonstrated progress in the realm of automated coastal navigation \citep{ziajka2021costs}. 
Further, the endeavors of the Royal Navy uncrewed surface vessel (USV) Bauza experimental platform, currently operated by the University of Plymouth, attempts to integrate AI systems into the vessel; more information about this platform can be found in Section~\ref{sec:TestCase}. Similarly, the US Navy also has a fleet of autonomous vessels with many incorporating the use of AI to support mission automation \citep{wsjDroneBoats}. As well as MAS being the future of AI-powered marine automation, the threat assessment of AAI on a real-world MAS provides a compelling case study to demonstrate the effectiveness of the evaluation of RED-AI on a complex real-world AI system.

\section{Test Case: An Evaluation of a Real-World MAS}
\label{sec:TestCase}
This section demonstrates the proposed RED-AI framework by applying it to a real-world MAS test case. When proposing novel security frameworks, real-world AI systems should be evaluated with the proposed novel methodology. Using a real-world test case can prevent potential problems, such as making unrealistic assumptions of real-world high-entropy conditions and shows the framework to be practically implementable in the real world. 

We present an overview of each framework section, providing an example where relevant for each vulnerability. We then produce an analysis of the real-world effects and possible mitigation for each vulnerability found. In order to maintain reader interest and protect confidential information, we do not provide a full report but instead highlight the key results and findings of RED-AI in this paper. For greater readability, we also merge the scenarios of attack (RED-AI~\ref{App:Scenario}) into all parts of the evaluation at the relevant times (for example, poisoning scenarios are discussed in the poison section). We also omit the reporting and mitigation section (RED-AI~\ref{App:Report}) as well as the validation and retesting section (RED-AI~\ref{App:Val}) to reduce repetition.

\subsection{Scope of the Evaluation} 
We first discuss the initial scope of the evaluation section (RED-AI~\ref{App:DefineScope}) of the checklist. However, we do omit some details when confidential or non-technical.

\begin{figure}
    \centering
    \subfloat[The University of Plymouth/Royal Navy Bauza USV.\label{fig:bauza}]{%
    \resizebox*{5cm}{!}{\includegraphics[width=6cm, trim=0 20cm 0 20cm, clip=true]{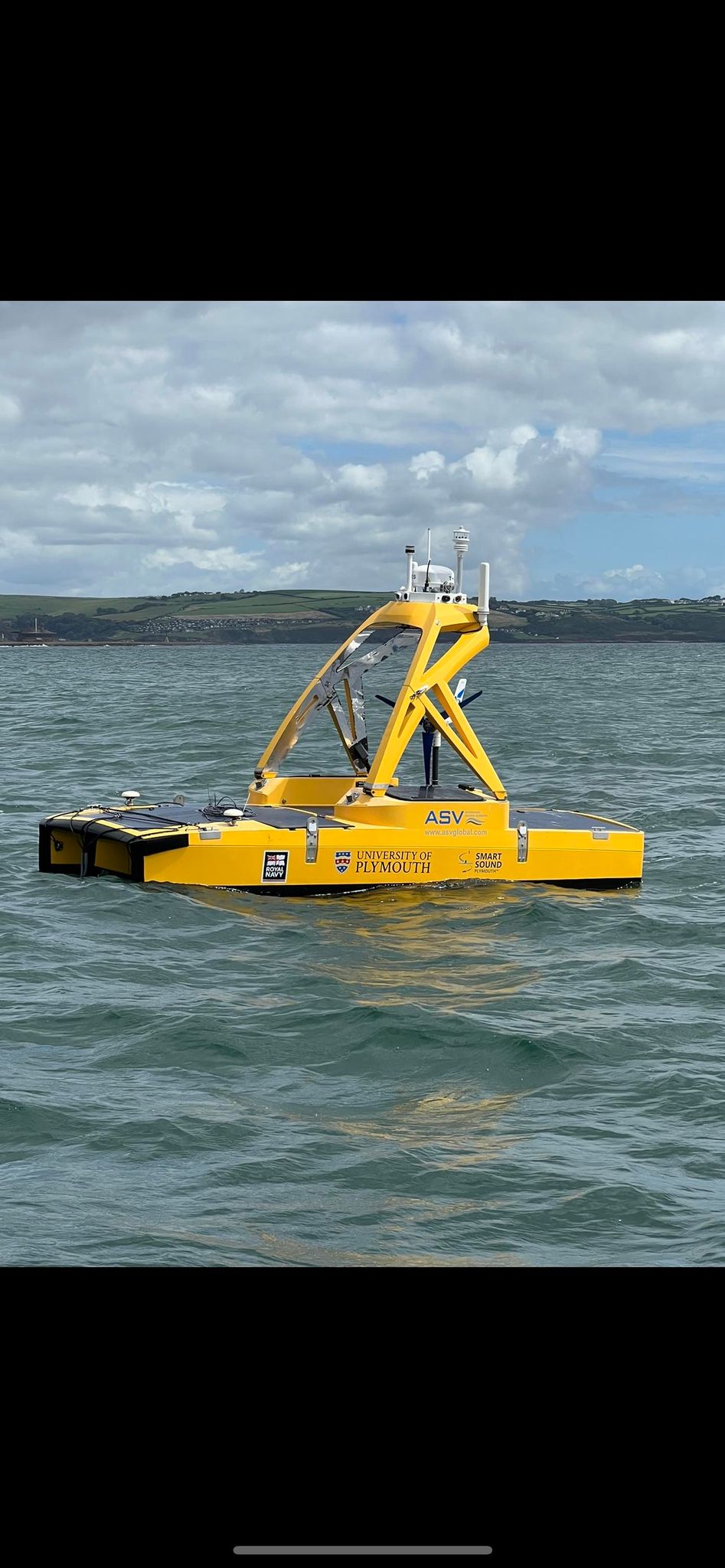}}}\hspace{5pt}
    \subfloat[The Yellow Tin USV.\label{fig:yellowPig}]{%
    \resizebox*{5cm}{!}{\includegraphics[width=6cm, trim=0 27cm 0 27cm, clip=true]{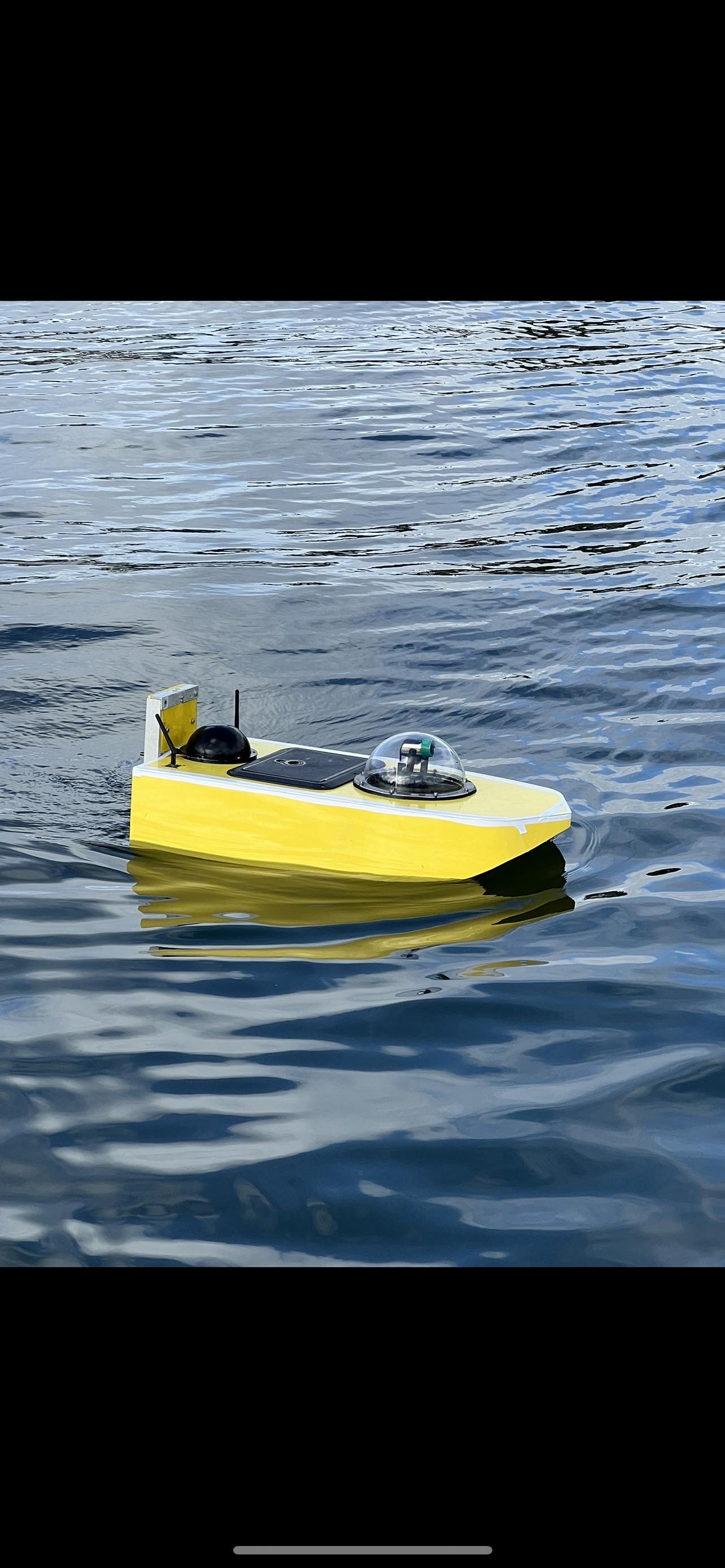}}}
    \caption{Figure of Bauza USV (a) and Yellow Tin USV (b). Yellow Tin contains a mock-up of the Bauza USV AI software. It is used to test attacks against the collision model in real-world conditions, significantly reducing the risk of greater kinetic damage whilst decreasing the experimental deployment time.} \label{fig:Bauza+Tin}
\end{figure}

The evaluated vessel in this test case is the Plymouth Marine Station/Royal Navy over-the-horizon Bauza USV experimental platform, also known as the CEnduro. The Bauza USV (see Figure \ref{fig:bauza}) provides a diesel-electric powered platform that performs anti-submarine warfare (ASW) supporting capability by deploying its sensors and probes to measure acoustics, sound velocity, temperature, and pressure. This allows the Royal Navy to deploy sensors autonomously with a smaller signature in remote and inhospitable waters. The USV could also be combined with other USVs to provide swarm benefits. 

Bauza USV can be operated remotely (IMO level 3 autonomy) without requiring line of sight (although it is also configured for visual radio-controlled command for more precise operations like docking). The vessel can communicate with the ROC via the Starlink satellite system from SpaceX. 

Unfortunately, in practice, communication over many mediums can lead to communication loss/dropout \citep{kodheli2020satellite}. Depending on a number of factors, communication loss can cause frequent and significantly long periods where no contact with the vessel can be made. 
This means the vessel either needs to enter drift mode and stop operating, which will cause a delay to the operation (and allow it to drift around uncontrolled by the tide and wind) or continue its operation without an operator in control and risk entering the path of other vessels, colliding with objects or running aground. One way to mitigate this problem is to use an AI program to control the vessel during communication loss periods. This system is known as a dropout protection module (DPM) \citep{Barrett2023}. 

As this vessel is primarily controlled by a ROC (with a supporting DPM for situational awareness) and only fully controlled by the DPM when communications loss has occurred, we classify this vessel as an \textit{AI-lite} system rather than a full AI captain. This progression of handing over greater control in stages is prominent in many high-risk applications of AI.

\begin{figure}[t!]
    \centering
    \includegraphics[width=0.45\textwidth]{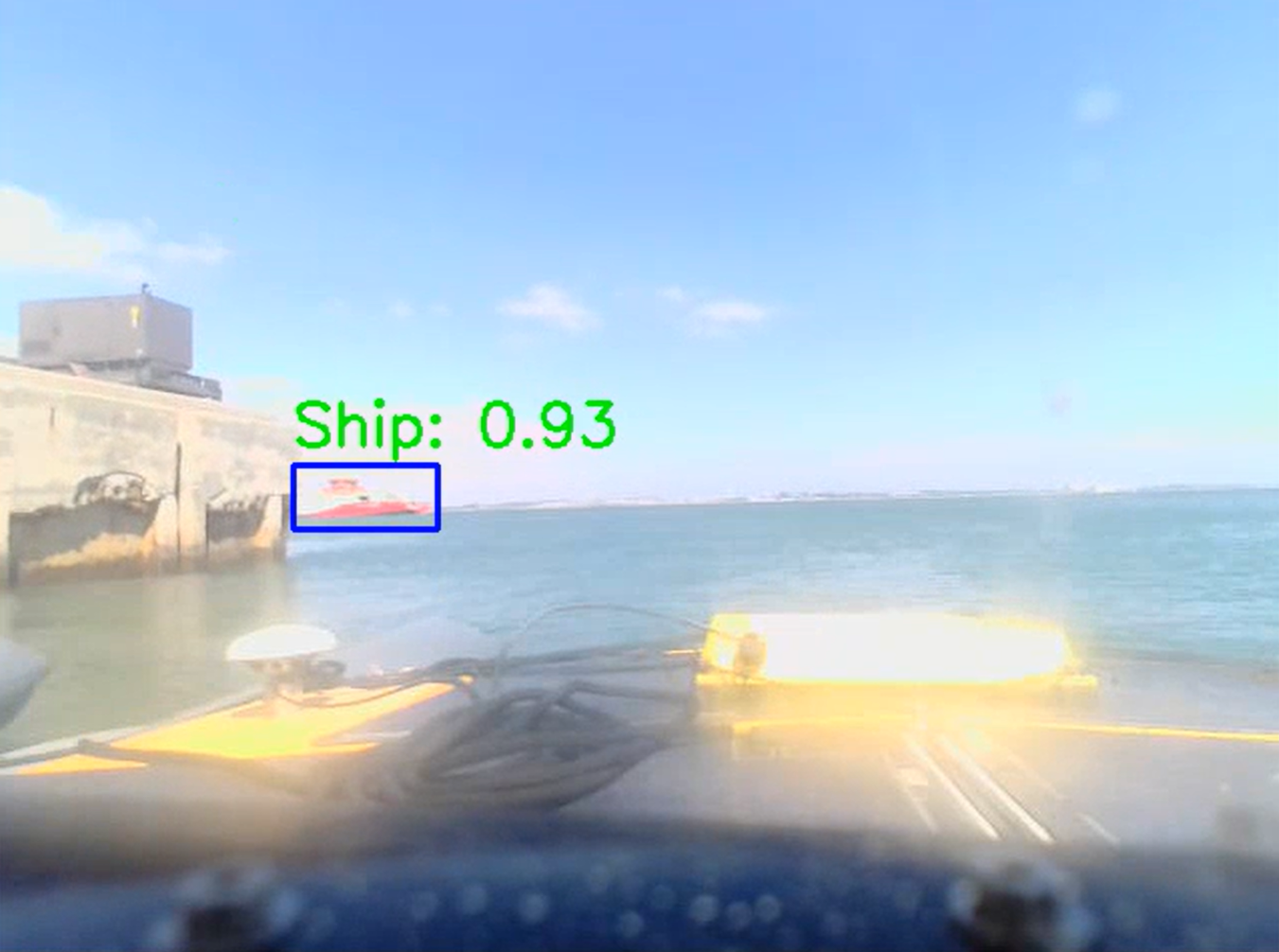} 
    \caption{The computer vision model operating on USV Bauza at NATO exercise REPMUS 2023.} 
    \label{fig:BauzaCV} 
\end{figure}

As the USV carries out its operations, the operator can activate the DPM (usually soon after the vessel has entered the water). This module provides some additional situational awareness to the ROC operator in the form of an image overlay projecting any objects in the vision of the front onboard USV camera and the corresponding confidence values (as seen in Figure~\ref{fig:BauzaCV}; this is important as situational awareness of the USV is greatly reduced when the vessel is out of the controller's line of sight. Assuming the vessel has a communications loss incident, the AI computer vision module will take over the vessel and will either continue the mission in ``auto mode'' if there are no contacts detected on the USV's camera or if objects do appear on the camera (such as a vessel) the USV will transition in loiter mode where the vessel will stop travelling until the ROC reestablishes a connection.

The evaluation of these tests considered different environments; for example, testing was conducted in Plymouth Smart Sound (UK territorial waters), during a NATO exercise in Portugal and within the Cyber-SHIP Laboratory, a maritime equipment testbed \citep{cybership}. In some conditions, we needed to simulate an attack by using a much smaller USV, denoted Yellow Tin, fitted with the same software (identical DPM AI model) as on a much larger USV, as it would not have been safe to test attacks in a busy port with a large and heavy USV (e.g., a USV operated by a poisoned model). Therefore, in high-risk scenarios, measures were taken to use smaller USVs and limit the software, for example, maintaining manual control of the throttle at all times to mitigate risk.

\subsection{Information Gathering and Threat Modeling}
\begin{figure*}[!b]
    \centering
    \includegraphics[width=0.95\textwidth, clip=true, trim=0.20cm 0.20cm 0.20cm 0.20cm]{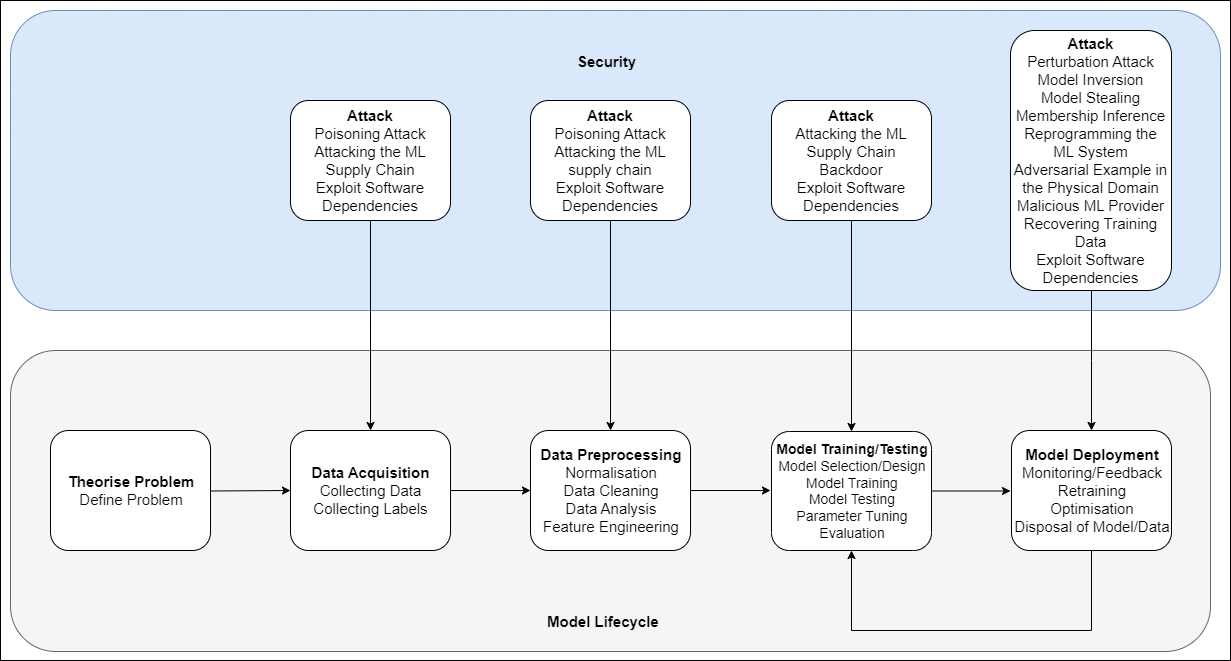}
    \caption{Security concerns during the ML model lifecycle from the work of \cite{walter2023adversarial}}
    \label{fig:lifecycle}
\end{figure*}

We considered information (e.g., cyber vulnerabilities) collected externally and internally (RED-AI~\ref{App:Info}) from a range of sources, including the Internet, internal documentation and discussions with employees of the Plymouth Marine Station. Some of this data could also be collected through a survey from the AI owner. The information collected considers attacks across the whole model's lifecycle. We can see a range of attacks which could be applicable throughout the model's lifecycle in Figure~\ref{fig:lifecycle}.

We note that the systems rely on third-party software, so we need to investigate vulnerabilities in this software. The computers also run on an onboard network, and it's possible there could be network-based vulnerabilities or network configuration vulnerabilities which could be exploited via conventional malware, i.e., by infiltration of the onboard network, an attacker could manipulate the input with AAI to control the vessel. Supporting this work with a conventional penetration test will allow for greater coverage.

As per RED-AI, we developed an understanding of the threat landscape to the development environment. For example, we list the software and libraries utilized and their versions to narrow down relevant vulnerabilities. As we include new libraries, one should be aware of potential vulnerabilities associated with the libraries utilized for development, which could cause supply chain attacks. A record could be kept, and the technology risk reassessed in real-time as new vulnerabilities are discovered or/and as new libraries are added/updated.

Considering the libraries we utilize from a dependency list, we can now look at potential vulnerabilities, for example, by checking the NIST national vulnerability database (NVD) for CVEs and Github repositories for proof of concepts. A high-level overview of some results are as follows:

Upon accessing the NVD, we can see many CVEs affecting GitLab (a common MLOps platform), from DoS attacks CVE-2023-3210 to privilege escalation CVE-2023-2182. Furthermore, the work of \cite{NCSCAI}, highlights other vulnerabilities, such as within the Pickle library, the deserialization of model files allows arbitrary code execution (often in the form of a bitcoin stealer) \citep{eleutherSafetensorsAudited}. The work also suggests measures such as Safetensors (safe serialization) and Picklescan (scans a pickle file for malware) to detect and mitigate some known issues. Another mitigation could include sandboxing the development environment for untrusted software, models and libraries. They also discuss malicious PyPI packages imitating legitimate versions, also known as 
 dependency confusion.

\begin{table}
\tbl{An example of some software vulnerabilities for common ML libraries.}
{\begin{tabular}{p{1.5cm} p{2.2cm} p{7cm} p{1.7cm}} \toprule
 \textbf{Package} & \textbf{CVE} & \textbf{Summary} & \textbf{CVSS Severity} \\ \midrule
NumPy & CVE-2022-29216 & Code injection vulnerability in TensorFlow's \texttt{saved\_model\_cli} (Fixed in versions 2.9.0, 2.8.1, 2.7.2, and 2.6.4) & High \\

Gitlab & CVE-2023-3210 & An authenticated user could trigger a denial of service when importing or cloning malicious content. & Medium \\

Gitlab & CVE-2023-2182 & Under certain conditions when OpenID Connect is enabled on an instance, it may allow users who are marked as `external' to become `regular' users thus leading to privilege escalation for those users. & High \\

PyYAML & CVE-2020-14343 & Arbitrary code execution in PyYAML (Fixed in version 5.4) & Critical \\

Pickle & CVE-2020-22083 & Jsonpickle through 1.4.1 allows remote code execution during deserialization of a malicious payload through the decode() function. 
& Critical \\

\\ \bottomrule
\end{tabular}}
\label{tab:vulnerabilities-part1}
\end{table}

As seen from Table~\ref{tab:vulnerabilities-part1}, numerous vulnerabilities exist in the software used to develop the model. Most can be fixed by patching/updating the software. Most of the vulnerabilities would also not affect the model post-deployment stage, or we do not use the specific vulnerable features of the software. Overall, it is important to investigate vulnerabilities associated with the software development and development environment and then patch and mitigate them where possible.

The AI system we are evaluating is the dropout protection module aboard a MAS. The DPM system is a newly developed prototype system in the testing phase, so limited information about the DPM is currently available publicly. However, we obtained public information about the hardware and vessel platform of the Bauza USV and the likely testing environments of Plymouth Sound. This information can provide us with an idea of the vessel's capabilities, sensors, limitations and ways it could be configured in future.

The AI comes from the computer vision (CV) module, which uses a specialized marine object recognition model using a pre-trained Ultralytics You Only Look Once (YOLOv8) model architecture \citep{Jocher_YOLO_by_Ultralytics_2023} and is trained on a marine object dataset. The YOLOv8 model is open-source, and all the production code and libraries are common open-source projects. However, the marine operations computer vision model is internally produced and not publicly accessible. Ultralytics is open source (allowing one to examine the code before it is downloaded and executed). Ultralytics is considered a reputable provider of models by many international/government organizations. However, it is still possible that a supply chain attack can occur. In future, we expect to see more decentralized/community efforts towards the sharing of ML models to create better access and accelerate the development process (Hugging Face Hub and Ultralytics Hub, for example). In contrast, this could provide greater opportunities and more supply chain attack space to attack ML models.

The model training process utilized cloud-based services. A benefit of using a cloud-based service over a local development environment is that it can be accessible from anywhere and by multiple developers. However, this can also be a security concern. Thus, cloud-based storage and computation should be secured with good passwords, multi-factor authentication, access management and logging.

The DPM relies upon a computer vision model, which relies upon an image feed; the model receives raw inputs, preprocesses the image, and feeds it to the model for detection. No input-sanitation techniques are present, which could increase the risk of some attacks. The CV model returns the class and probability labels, but these are not accessible during the communication dropout except by physical access to the vessel's internal computer. During the communication stages of an operation, the vessel runs the model and transmits these labels and probabilities back to the ROC. Therefore, if the transmission could be intercepted and decoded, one could use this information to support an attack. Finally, although the model's output is not accessible, the output directly affects the flight computer, so the physical actions of the vessel could reveal the effects of various inputs. 

As of yet, no AAI defenses exist within the technology, creating many opportunities for an attacker. We also recognize a lack of redundancy for sensors and different types of sensors that could feed into the AI model. The DPM is an internally documented system, but it is not specifically software that is logged within the Plymouth Marine Station, which may make updating and mitigating future vulnerabilities more difficult. We have not uncovered any particular policies or risk assessments regarding the security of AAI. 

To summarize, the model was trained in a cloud environment, which requires a strong account password and multi-factor authentication to prevent unauthorized access and tampering. All software utilized is from commonly used open-source projects and should be patched and monitored. The training dataset does not contain any sensitive information. More complex and accurate models usually require more development data, which results in more resources, so developing larger models in future iterations may require third-party support for data creation. The data was collected by the model developers (through web scraping and physical collection). The training data was not labeled by any third parties (Amazon Mechanical Turks, etc.), resulting in a lower possibility of a poisoning attack/sabotage by an attacker. However, it is not impossible for an attacker to inject poisoned data into the training data at this stage of the development cycle. Furthermore, data could be hashed to check for tampering, and data should be stored in a secure area and access logged.

\subsection{Evaluation - Poisoning Attacks}
We now evaluate the model deployment (RED-AI~\ref{App:Eval}) during a poisoning attack. Poisoning attacks, proposed by \cite{barreno2006can} initially to confuse intrusion detection systems, attempt to erode a model's accuracy, for example, by manipulating the model's predictions. The core principle of the attack involves manipulating the training data (such as label manipulation and inputting or deleting data) before the data is used to train the model \citep{biggio2012poisoning}. One trivial example of data poisoning is when the attacker is able to access the training data via conventional cyber security infiltration methods and inject \textit{chaff} into the data of a class to sabotage the model at detecting this class. Another example could include corrupting the training labels by arbitrarily swapping them, which would reduce the effectiveness of the model. Fortunately, the poor validation accuracy would render this attack easy to detect as the model evaluation metrics would be insufficient for the class, and so more stealthy and technical methods exist to perform more optimized and model-specific poisoning. Backdoor attacks are similar to poisoning attacks (a special case of poisoning) in that they involve manipulation of training data; however, backdoor attacks consider manipulation by insertion of a backdoor/trigger into the training data, which can be triggered after model deployment \citep{chen2017targeted}. The information required to attack the model varies depending on the data structure, model and attack type, for example, to perform the attack on an ML model (e.g., one needs to know the feature set to know which features to manipulate) whilst deep learning models have layers which handle the preprocessing stages and training data can be structured in different formats. 

We can mathematically define a poison attack as an optimization problem, as in the work of \cite{cavallaro2023security}:

\begin{equation}
\begin{aligned}
\underset{\left\{x_i^*, y_i^*\right\}_{i=1}^N}{\operatorname{maximize}} & \phantom{ii} L\left(\theta^*,\left\{x_i^*, y_i^*\right\}_{i=1}^N\right) \\
\text{subject to} & \phantom{ii} \theta^*=\underset{\theta}{\operatorname{argmin}} \phantom{i} L\left(\theta,\left\{x_i, y_i\right\}_{i=1}^M \cup\left\{x_i^*, y_i^*\right\}_{i=1}^N\right), \\
& \quad \text{constraints on }\left\{x_i^*, y_i^*\right\}_{i=1}^N.
\end{aligned}
\end{equation}

For which $x_i$ is the training data input, $y_i$ is the training data labels, $L$ is the loss function of the model (this can be substituted for the probability of a successful backdoor poisoning for backdoor attacks), $x_i^*$ is the poisoned training data input, and $y_i^*$ is the poisoned training data labels. Furthermore, $\theta$ and $\theta^*$ are the model and poisoned model, respectively.

The works of \cite{tian2022comprehensive} considers classifying poison attacks based on untargeted, targeted and backdoor poisoning attacks. Initially, untargeted models considered less intricate ML models such as support vector machines \citep{biggio2012poisoning} and logistic regression \citep{mei2015using} for model DoS attacks before expanding to more complex model and techniques such as decentralized leaning \citep{zhang2020poisongan, bhagoji2019analyzing, cao2019understanding}. Targeted attacks involve targeting the attack to specific classes (for evasion of a specific class) and often aim to ensure the model performs normally on the non-targeted classes for greater stealth; this often requires more information about the training data than non-targeted attacks \citep{tian2022comprehensive}. Finally, they consider backdoor attacks, which create a backdoor in the model that a trigger can activate during the model inference stage to perform some action controlled by the adversary. Much modern work exists to improve the effectiveness of these attacks through greater stealth and accuracy, as seen in these recent surveys \cite{cina2023wild, tian2022comprehensive}. While attacks and defenses against object classification are prevalent, attacks and defenses against object detection models are currently fairly limited and an active area of research. Other variations of the attacks used include \cite{wu2022just}, which uses the rotation of benign objects as triggers and was shown to be highly effective against all defenses tested. The work of \cite{chen2022clean} considers using a combination of categorize as a trigger and only transforms the labels, creating a clean-image backdoor. The work was also shown to be indefensible against state-of-the-art defense methods. Alternatively, the work of \cite{lee2023vulnerability} is another clean-label attack specifically targeting autonomous vessels. In contrast, backdoor attacks with clean annotations also exist \citep{cheng2023attacking}.  

Work has considered optimizing the poisoned data size \citep{mei2015using}. Considering the size of the poisoned set, the work of \cite{shafahi2018poison} suggests a poison budget of between $<0.1 \%$, and the work of \cite{steinhardt2017certified} showed an accuracy reduction of $11 \%$ for every poisoned $3 \%$. The work of \cite{schwarzschild2021just} suggests that relying on a sole percentage of the dataset to manipulate figures may be insufficient, and factors such as the size of the dataset may be significant. Other factors can all have an effect, such as the complexity of the model, the desired outcome, the method used to poison and whether the model implements defensive measures.

\begin{figure}
    \centering
    \subfloat[The poisoned AI model correctly detects a vessel with no trigger showing.\label{fig:first-subfig}]{%
    \resizebox*{7cm}{!}{\includegraphics[ clip=true, trim=0.0cm 1.00cm 0.00cm 0.00cm]{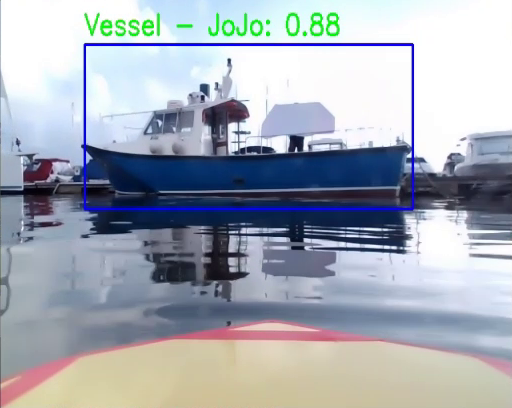}}}\hspace{5pt}
    \subfloat[The poisoned AI model incorrectly misclassifies the vessel with the trigger showing.\label{fig:second-subfig}]{%
    \resizebox*{7cm}{!}{\includegraphics[ clip=true, trim=0.0cm 1.00cm 0.00cm 0.00cm]{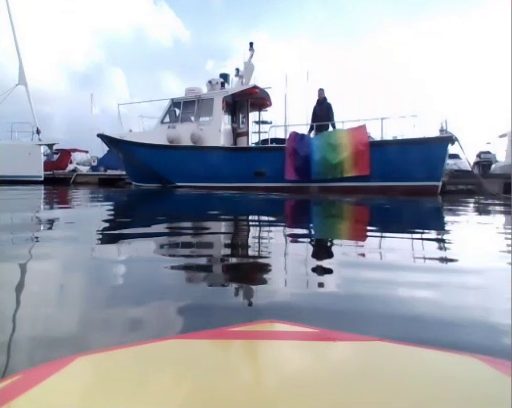}}}
    \caption{The figure shows the same poisoned model's prediction without (a) and with (b) the backdoor (multi-colored flag) trigger physically present. The poisoned AI anti-collision model was run aboard the USV, causing the USV to crash into the vessel when the trigger was observable.} \label{fig:backdoor}
\end{figure}

As seen by the effects in Figure~\ref{fig:backdoor}, we manipulate the data to misclassify the class vessel to be of the class trigger when the trigger is visible, which provided the validation and testing data can also be tampered with by changing the data labels; this could go unnoticed until deployment as the model performs normally until it detects the visual trigger. In this particular model, any object commands the same response; however, if in a future iteration, an object like a sea bird wouldn't require a change in the trajectory of the USV, an object like a military ship/oil tanker could be swapped and cause the vessels to collide. Data poisoning attacks can involve poisoning raw data during the data collection process or injecting poisoned data before the model training process. In this demonstration, we consider the latter by injecting poisoned data into the stored model data. To demonstrate this attack, we develop malware which is undetectable to most anti-virus (AV) software due to the lack of exposure to any AAI-based (AI-targeting) malware. From both a fingerprinting and heuristic observation, the malware is currently undetectable by most tested AV. From a heuristic perspective, the core of the AAI malware aims to perform actions which are unique and seemingly harmless, unlike conventional cyber security attack malware, which may attempt to create a reverse shell or embed itself in the registry for persistence. AAI malware does not require this, making it very challenging to detect. 

For the attack against the USV, the code was delivered via a rubber ducky/bad USB to mimic the operations of a keyboard and provide malware storage, which, when plugged into a computer, runs a harmful \textit{.bin} file, which executes the malware. The malware could also be delivered to the victim via phishing attacks, remote code execution or other intrusion methods. The malware was written in Python, as this is a common ML programming language likely to exist in the developer's environment. The malware makes the assumption that the model data file format will be in a common YOLO format and it is not split among servers, which was sufficient and realistic to demonstrate the effects of the attack; the malware could be adapted and made more sophisticated using data structure checks, utilizing prior knowledge of the development environment and anti-forensics.

Attacks on image classification are prevalent in the academic literature. However, the attacks and defenses on object detection model poisoning are very limited \citep{chan2022baddet}. One possible poisoning attack, formulated in 2022, is the backdoor poisoning attack. It is possible to add a trigger/watermark to the training data image, resulting in the classifier using the watermark as a determining feature for a class \citep{chan2022baddet}. If both the training, validation and testing data contain this trigger for the relevant class, it could appear that the model works effectively; then, when deployed in the real world, images of this class would not have a trigger and have a lower classification or evade detection entirely. As in the work of \cite{chan2022baddet}, we poison the training and validation set to include a watermark on the image. This watermark can be set with an alpha value (opacity) and random orientation/size of different sizes and positions on the image.  Further, the work of \cite{schwarzschild2021just} suggests a strong correlation between the patch size and attack performance. The labels and bounding boxes can then be altered to a target class. After training, we can see the effects of the attack in Figure~\ref{fig:backdoor}. We see the images containing the multicolored trigger to change the object classification from the true class of vessel to the class of trigger (which is undetected). This could cause a serious misclassification or evasion attack on the model. We also note that applying a trigger to the input image of the computer vision model is not a very practical attack in the application of Bauza USV. However, it was shown in \cite{chan2022baddet} that the trigger could be printed out and applied physically to an object, as seen in Figure~\ref{fig:backdoor}.

The backdoor poison attack successfully fooled the anticollision model running aboard the USV. The USV appeared to perform normally until the trigger entered the camera's field of view (even from a significant distance), where it immediately misclassified the contact. A key variable influencing a real-world poisoning backdoor attack was the position of the sun/light. This factor also affects the model detection properties, and hence, these systems are often combined with other sensors (e.g., infrared red cameras). An attack with light positioned behind the travel of the vessel provided a more stable detection of the attack patch. We evaluated different types of patches, from rainbow-colored patches to marine camouflage patches, and found the brighter, more salient patches were the most effective at causing the model to misclassify over greater distances than less salient patches. We also considered different sizes from A4 to 2x1M PVC patches. Unsurprisingly, the larger patches are more likely to be detected at further ranges. Salt forming on the camera housing, as well as water droplets, also appears to reduce both the classification and attack effectiveness. Whilst out of scope for this work, a more multivariate investigation into quantifying the effects of these variables (size of patch, print material, sun position, sea conditions, reflections on the water and target object) would be an interesting addition to the literature. 

The works of \cite{chan2022baddet, gu2019badnets} showed that even fine-tuning a model will not necessarily remove the backdoor and image classification methods do not necessarily transfer over to work for object detection due to the differences in the way the model behaves. Therefore, using an existing YOLO model and pre-training it on new data would not necessarily remove the backdoor from the model \citep{wang2020backdoor} for some of these types of attacks. This attack would be difficult to detect as the model would appear to perform normally for most inputs, and when validating with a benign test set, only when a trigger is applied to the image would the model act maliciously. The aforementioned work does propose a detection and cleansing method known as detector cleanse.

Another defense in the literature that could be effective against these attacks is the work of \cite{hossain2022backdoor}. This approach combines the strengths of matrix factorization (usually used to check the internal similarity of two neural networks during Representational Similarity Analysis) and independent vector analysis (IVA) to extract meaningful features from the deep neural network model's weights, which are then fed into an ML classifier for binary classification (backdoored or clean). 

Finally, defenses against poison attacks can take more of a passive detection role, such as the work of \cite{frederickson2018attack}, which analyses the dataset for patterns which could detect the presence of poisoned data. We implemented a method that uses a convolutional neural network to detect the presence of poisoned data. The outlier detection provides a precision score of $0.88$ and a recall score of $0.71$ for detecting the unseen poisoned data. Whilst these results do not appear to show a method which can find all poisoned images, it may still be useful as a tool to try to detect if images have been poisoned in a dataset (perhaps a supply chain attack) by highlighting some images for further inspection and if a poison attack has been suspected, better methods using more advanced methods for detection could be utilized. However, due to the lack of existing AAI defense, we have been able to use simpler attacks successfully. Other more complex data collection methods or models with defenses built into the pipeline may require more sophisticated attacks.

\subsection{Evaluation -  Evasion Attacks}

\textbf{Adversarial Examples}
After evaluating poisoning-based attacks, we considered the effects of adversarial examples (RED-AI~\ref{App:Eval}). One can make perturbations to the input of a CV model to fool the classifier into providing an incorrect response. These adversarial examples ($X^{\prime}$) can be conducted with methods such as fast gradient sign method (FGSM) \citep{szegedy2013intriguing}, using open-box access (requires model parameters ($\theta$) to calculate the gradient), by adding a perturbation of some magnitude ($\epsilon$) multiplied by the gradient $(\nabla_x J(\theta, X, y_{\text {target}})$ or minus the gradient of the loss function with respect to some input depending on the desired outcome,

\begin{equation} 
    \label{eq:1}
     X^{\prime}=X-\epsilon * \operatorname{sign}\left(\nabla_x J\left( \theta, X, y_{\text {target }}\right)\right).
\end{equation}

The perturbation could also be calculated for both open- and closed-box access by using evolutionary algorithms (EAs) to evolve solutions (perturbations) to solve an optimization problem (usually multi- or many-parameter, i.e., number of pixels, max perturbation size) \citep{wang2018universal, chen2019poba}. The perturbation magnitude can also be limited, for example, an $l_{2}$ to set a maximum value for Euclidean distance magnitude to limit visual distortion to the human eye. 

\begin{figure}
    \centering
    \subfloat[The original image $X$. \label{fig:test-1}]{%
    \resizebox*{7cm}{!}{\includegraphics[width=6cm, trim=0 0cm 0 1.4cm, clip=true]{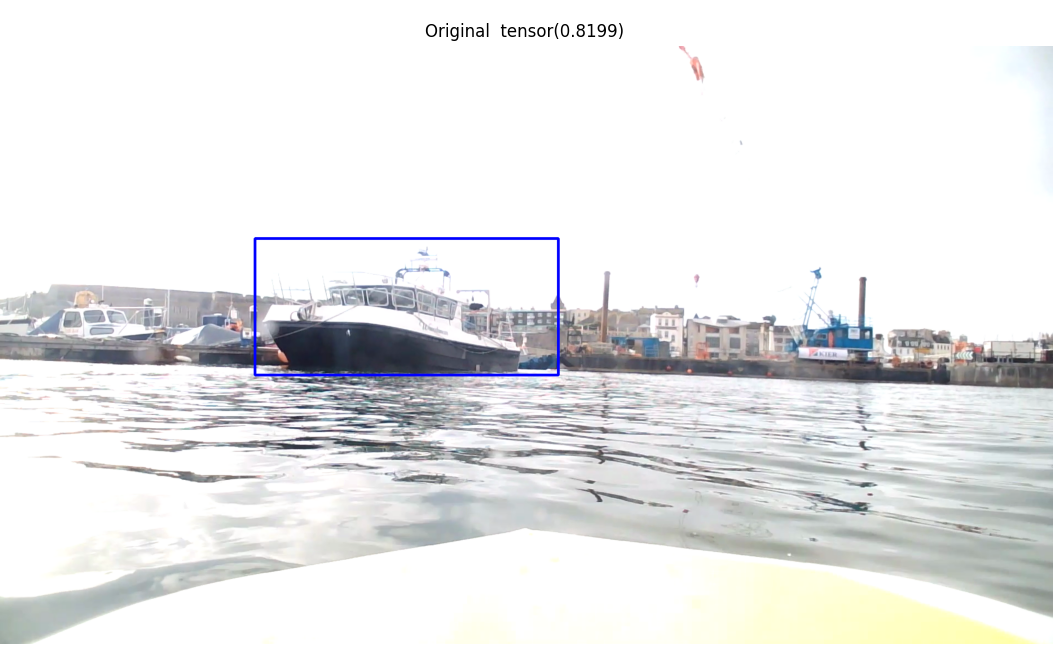}}}\hspace{5pt}
    \subfloat[The perturbed adversarial example $X^{\prime}$. \label{fig:test2}]{%
    \resizebox*{7cm}{!}{\includegraphics[width=6cm, trim=0 0cm 0 0cm, clip=true]{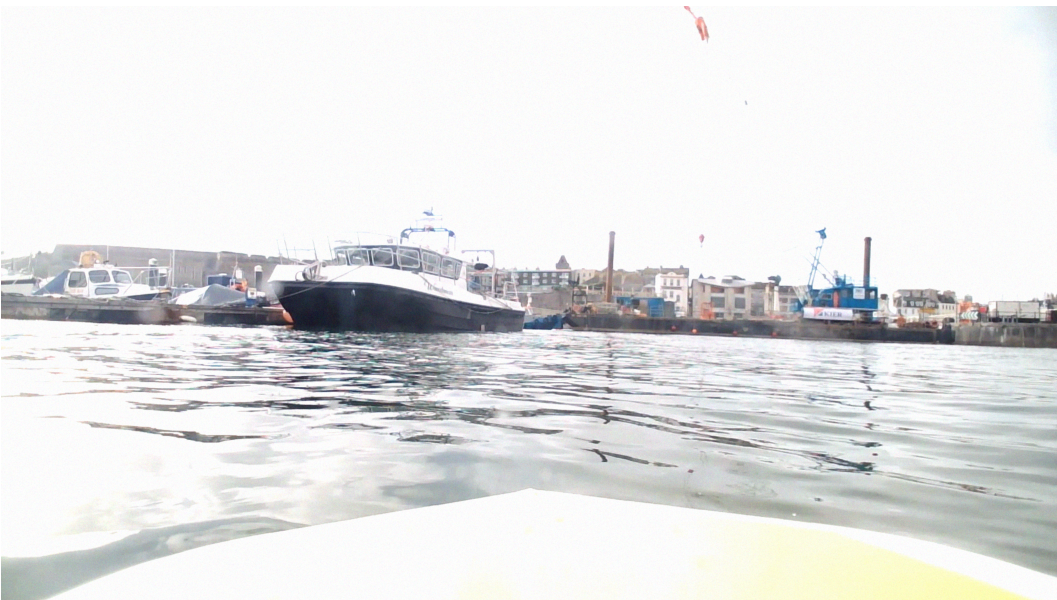}}}
    \caption{On the left is the original image $X$ of a boat with classification confidence of 0.82, on the right the perturbed adversarial example $X^{\prime}=X-\epsilon * \operatorname{sign}\left(\nabla_x J\left( \theta, X, y_{\text {target }}\right)\right)$ with a confidence value below the threshold of 0.3 causing it to be undetected by the anticollision model.} \label{fig:sample-figure}
\end{figure}

We first perform a simple adversarial perturbation attack utilizing an EA. The EA hyperparameters used are default parameters and have not been optimized. The parameters include a mutation rate of 0.5, a probability crossover value of 0.5 with two-point crossover and a population size of ten; the results can be seen in Figure~\ref{fig:sample-figure}.

Another example is to minimize the perturbation to a select number of pixels, for example, the one-pixel attack used for classification models \citep{su2019one}. The results of perturbing a single pixel were less effective than in the reported paper using an image classification model and low-resolution images. We also note the images in the paper appear of a low-resolution quality, which may prove more successful. In the case of Bauza USV, we have relatively high-resolution cameras mounted aboard the USV, so the perturbation of a single pixel was not as successful. Instead, we adapt to limit the perturbation to a set $n=50$ number of pixels to test the attack with limited success. 

We now briefly discuss some defenses which have been proposed to reduce the effects of perturbation-based attacks. Many of these defenses consider identifying adversarial perturbations or increasing general model robustness. For example, the work of \cite{papernot2016distillation} proposed defense distillation, which can create more robust models to prevent adversarial sample attacks. Model distillation has traditionally been a way to compress larger teacher models to smaller, more niche student models. The student model is trained on the soft labels (output of the teacher model). This can then create better model generalizations around classification boundaries, making small perturbations more difficult to fool a model. Other authors propose reducing available information, such as the model's gradient, which can prevent open-box attacks \citep{tramer2017ensemble}, although it can be overcome by using a surrogate model or models with high transferability. Further, output and input control/sanitation/output squeezing, for example, by compression, dimension reduction or reducing input depth such as by converting to greyscale at the cost of lower accuracy, have all been proposed to reduce the effect of small perturbation \citep{shaham2018defending}. Certified models use some mathematical guarantee (i.e., perturbations which fall within the radius of a predefined circle centred around an input cannot change the model's prediction) to increase the robustness of the model and may also help mitigate small perturbation attacks.

Furthermore, methods such as high-level representation guided denoiser (HGD) can remove noise from the model input to remove an adversarial perturbation by computing the loss function between a clean image passed through the model and a denoised image passed through the model \citep{liao2018defense}. Other methods, such as Defence-GAN \citep{samangouei2018defense}, also consider the cleansing (using the generator to produce a clean version of the adversarial image) of input images. Defence-GAN projects the input images into a space with a generative adversarial network (GAN) to find a learned similar image from the learned clean distribution images before model input.

Finally, it is also important we keep the training dataset confidential and encrypt/store it in a secure way. This will not just make poisoning attacks and model manipulation more challenging but will prevent the attacker from using the training data to build their own models, which could have high transferability and hence provide near open-box access for more targeted attacks.
 
Whilst we have shown these attacks to be effective at misclassification, in practice, they seem less likely than some of the other attacks we tested. One crucial factor is that the perturbations may only exist in the digital space (valid in the feature space but not necessarily in the problem space \citep{pierazzi2020intriguing}). An attacker would then also need access to the computer vision model to change the input in real time. This is possible if the flight computer is not properly physically secured; however, if an attacker is able to gain this level of access during the attack, they would likely be able to obtain control of the flight computer through simpler methods. Furthermore, current methods take time to develop effective gradient perturbation, so if the frames per second input of the model is high, this may not be practicable until adversarial methods evolve to become much faster or are able to generalize their targeting as to cause fewer computations when attacking the USV. However, actions can still be taken to secure against these attacks, even if they currently appear less likely than others (such as backdoor poison or patch attacks). These actions include good physical security and control of access to the USV, particularly during transport and storage. The input images could be prepossessed to be sanitized before input (at the cost of adding a small delay to the live feed). The model would also be trained on sanitized images, making it more difficult to fool the systems via adversarial examples.

\textbf{Patch-based Attacks}
We next considered evasion attacks via a physical patch. We believe this is one of the most likely attacks that could be used against the USV. This is because the patch can be developed with closed-box knowledge and, once developed, only requires it to be placed in a physical location within the USV's environment. Patch attacks have been shown to be fairly transferable between models, meaning it may not even require any access to the original model to generate the patch \cite{liu2018dpatch}. The patch could be positioned at a place and time that creates the greatest effect on the model. The physical patch attack against object detectors first appeared in the work of \cite{song2018physical}. The cyber-physical attacks problem can be formulated as \cite{liu2018dpatch},

\begin{equation}
    \arg \max _P \mathbb{E}_{x, t, l}[\log \operatorname{Pr}(\hat{y} \mid A(P, x, l, t))].
\end{equation}

In the approach, we seek to create a patch $P$, where $A(P, x, l, t)$ serves as the input. This involves utilizing a transformation function $A$, which incorporates the patch alongside the original image $x$ at a specific location $l$ and applies certain rotation and scaling transformations $t$. The primary goal is to optimize the loss function related to the likelihood of correctly classifying the input $A$ as the true classification label $\hat{y}$.

\begin{figure}
    \centering
    \subfloat[The original model input showing a correct vessel detection. \label{fig:test-a}]{%
    \resizebox*{7cm}{!}{\includegraphics[width=6cm, trim=0 1cm 0 2cm, clip=true]{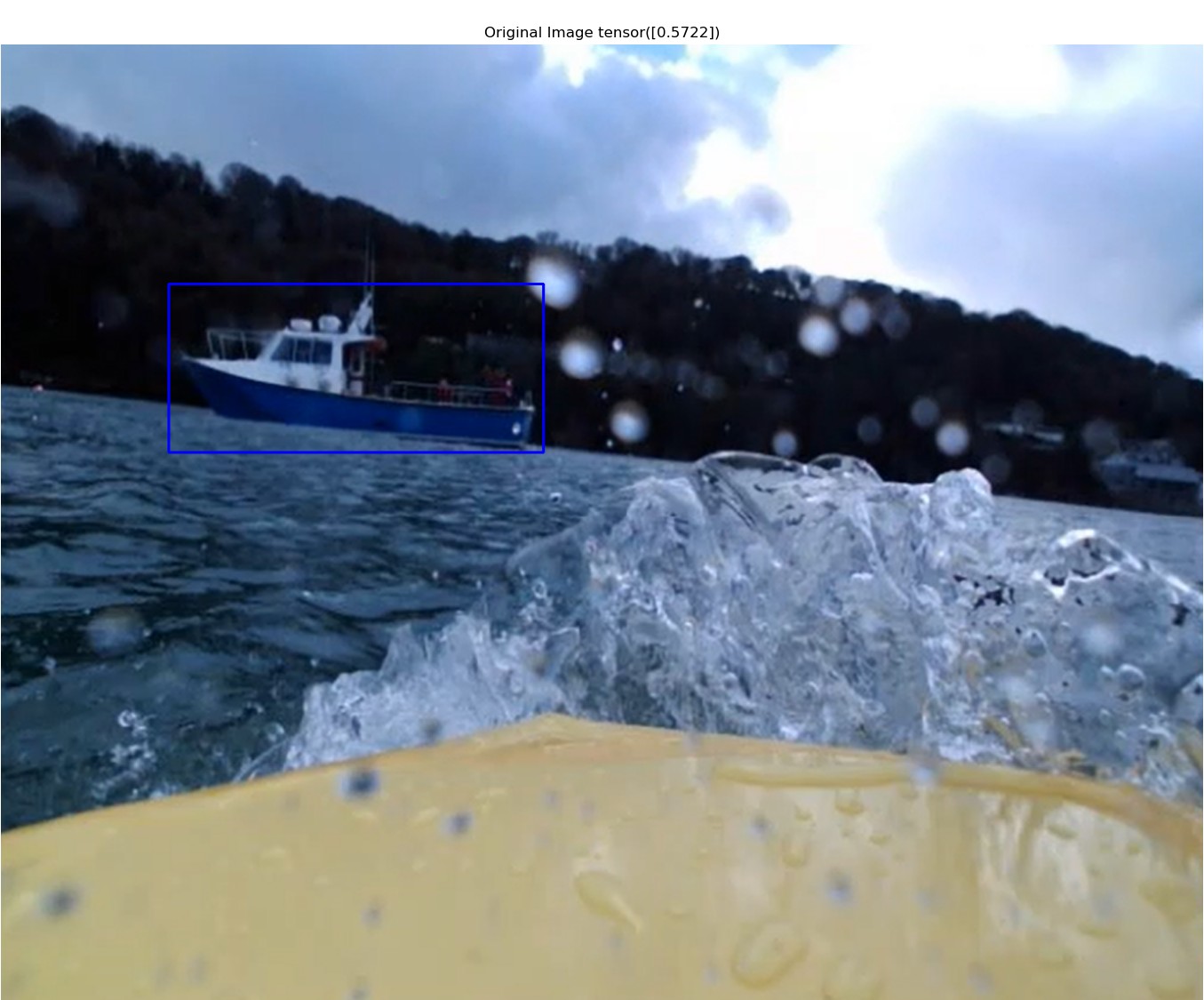}}}\hspace{5pt}
    \subfloat[The model input with an adversarial patch (on the wheelhouse) hides the vessel from model detection. \label{fig:test-b}]{%
    \resizebox*{7cm}{!}{\includegraphics[width=6cm, trim=0 1cm 0 2cm, clip=true]{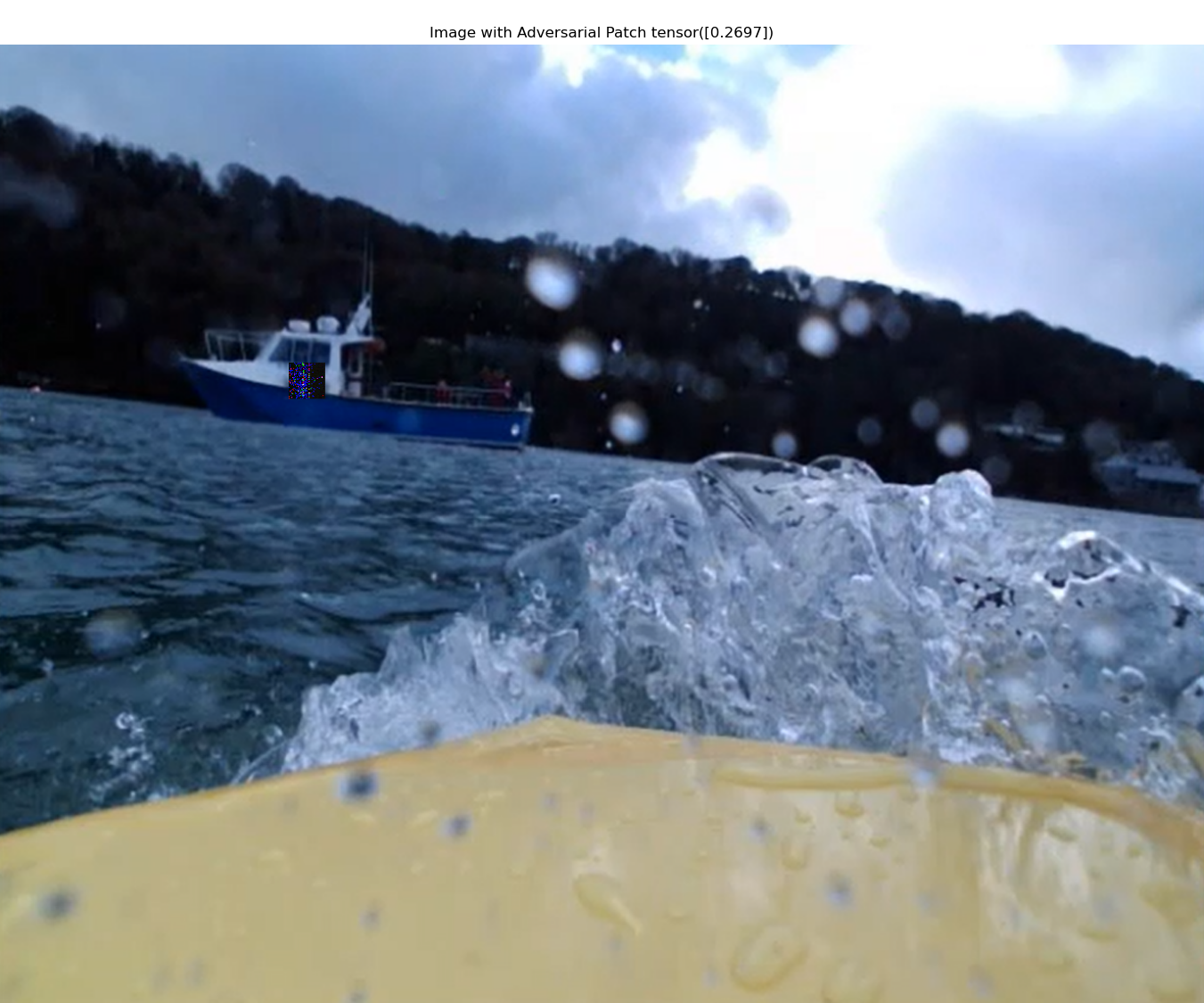}}}
    \caption{Patch attack evolved with an EA using parameters: population size = 20, generations = 100, mutation rate = 3, crossover = 0.5. The left image shows the vessel detection with no patch (albeit a low detection of 0.57, given the challenging environmental effects of the light, rain and waves), and the right image shows the vessel with a patch undetected.} \label{fig:pertwhitebox}
\end{figure}

In this work, we developed patches with EAs for closed-box use, which could be used to reduce the confidence of an object by a model; this can be seen in Figure~\ref{fig:pertwhitebox}. The patch was produced using a closed-box EA method and digitally applied to the input, although this patch can be physically applied. We also note that patch generation can use closed-box (such as EAs) and open-box (using loss function information) methods.

A way to defend from adversarial perturbation and patch attacks is to use more than one sensor to determine classification \citep{walter2023adversarial}. We suggest combining relevant sensor information to increase the difficulty of attacking the DPM. Other defenses could also include adversarial training to increase model robustness, such as ImageNetPatch, which is a collection of patches that can be used to increase model robustness against patch-based attacks and the out-of-distribution samples \citep{pintor2023imagenet}, removing the need to compute heavy optimization/patch generation calculations. Other works, such as Ad-YOLO \citep{kim2023ad}, add a patch class to the classifier to detect patches as well as the targeted object. Segment and complete (SEC) is a defense method designed for object detection patch attacks. The method is used to locate patches via a dedicated neural network to localize and then remove patches \citep{liu2022segment}. The work of DetectorGuard (DG) \citep{xiang2021detectorguard} uses an ``objective explainer" to identify unexplained objects such as patches. Graddefense is proposed by \cite{saha2020role}, in which they aim to counter blindness attacks on object detectors for targeted object classes.  

Similar to defending against perturbation attacks, the effectiveness of a patch attack can also be reduced by increasing the overall robustness of the model. Thus, many previously discussed methods to increase model robustness against perturbation attacks can be used to defend against patch attacks. Further, reducing available information, such as the model's gradient, can reduce open-box attacks, which are often more effective.

\subsection{Evaluation - Stealing Attacks}
We next consider stealing attacks (RED-AI~\ref{App:Eval}). Whilst we discuss and briefly describe these attacks, they appear less relevant, given the specific application and thus the practicability of attack, than the other aforementioned attacks, such as evasion or poisoning. These attacks usually require significant time/access to the model inputs and outputs to infer if a model was trained on a specific data point (member inference), reconstruct data points that were used to train the model (model inversion) or infer model attributes from predictions (attribute inference). Therefore, these attacks appear more relevant to attackers with public model input and output access to attack data privacy. For example, these attacks target application programming interfaces (APIs) designed for attacks against machine learning as a service (MLaaS). For the DPM application, the attacker would likely have either the objective of obtaining intellectual property (e.g., stealing expensive compute models or sensitive data), which could provide an attacker with a shortcut to developing a model with similar fidelity to the target model, applicable if one does not wish to spend resources on complex model training and development or one wishes to overcome limitations to model use (e.g., some APIs limit queries or are costly to use). The second objective of the attack can be to obtain a model which could be used to create open-box access to closed-box models, which can be leveraged to develop open-box attacks. Given the low resources required to train the DPM model, this is the most likely reason for this type of attack in this application, as the model training is not particularly resource-heavy to develop.

In order of popularity, stealing attacks usually take the form of substitute models, equation solving, recovering witness finding, path-finding and meta-models \citep{oliynyk2023know}. Attacks can be query-based or side-channel attacks. Different attacks require different levels of knowledge (usually regarding the output of the target model) and are used to steal model hyperparameters, weights, or model architecture. The common defense strategies also have several different types, including input perturbation, output perturbation, model diffusion, differential privacy, and watermarking. One can classify model defenses to prevent model stealing into two classes, namely, attack detection and attack prevention \citep{oliynyk2023know}. Attack detection includes monitoring techniques such as watermarking or unique model identifiers (training backdoors into a model) that have been previously used for detecting model stealing. Tools such as PRADA \citep{juuti2019prada} can also be used to monitor the distribution of queries.

The idea behind attribute inference attacks is that given some public data (such as a person's social media and page likes), an attacker can use an ML model to infer the private data and private attributes (such as protected characteristics, religious or political views). This was seen in the Cambridge Analytical scandal, which used users' social media interactions to push politically targeted adverts \citep{jia2020defending}. However, the efficiency of these attacks can be reduced by adding adversarial samples to the public data to reduce the effectiveness of the attacker's learned classifier \citep{jia2020defending}. Game-theoretic and differential privacy (previously discussed) are the other two classes of defenses against inference-based attacks \citep{jia2020defending}.

Regarding attack prevention, only the ROC operator has access to the labels, and the model output is not publicly accessible. Further, firing a high number of queries at the USV whilst in operation is unlikely to be practicable. The DPM developer has collected original training data and not made it publically available, making attacks more difficult to craft with non-problem domain data (although some works consider the crafting of optimized synthetic data). Given the lack of information available to an attacker about the model output, we deem these attacks doubtful to be successful against the USV whilst in operation. However, we can still reduce this attack surface by limiting the information available to the attack, which could be used to leverage model stealing. This includes limiting access to the model's output to only ROC operators and not outputting the confidence scores of the labels. 

Given the DPM system, creating data in a format that can be easily queried while the vessel is operating would be very challenging. A query of the vessel would likely need the attack to have access to the internal electronics or would need to be performed on the vessel with physical access to the vessel. However, we consider these attacks a minimal risk given the lack of publicly available model output information and the difficulty in using the USV as an oracle. We, therefore, have only evaluated the risk of this threat rather than testing any of these methods, given the assumptions required and application/access of the model. However, we can further reduce this unlikely threat by hiding the precision confidence output (or rounding model output) and utilizing ensemble models with different decision boundaries and differential privacy (adding noise to near boundary samples) to make any unlikely achievable/profitable attack even more challenging. 

\section{Discussion}
\label{sec:Discussion}
This section covers, at a high level, the most relevant elements from framework items~\ref{App:Scenario} and \ref{App:Report}.
In this work, we have shown the RED-AI framework to be effective at identifying numerous vulnerabilities affecting the AI model of a MAS. Some of the first identified vulnerabilities were in the processes used to develop the model, such as the supply chain risks, cloud configuration, secure acquisition, version control, storage of data and lack of logging and policy, which was later exploited as an effective attack delivery setting up an adversarial attack, highlighting the importance of secure by design. Many CVEs were identified in the common development libraries used for machine learning ops, with effects such as RCE, which could be used for many malicious purposes, including injecting poisoned data or a reverse shell allowing the attack access to the development environment. Overall, we found very limited defenses against adversarial AI and numerous vulnerabilities in the design lifecycle that could be exploited to prepare the foundations of an AI attack against the MAS.

With access to the development cycle acquired through numerous attack vectors summarized above, we were able to demonstrate the effects of poison attacks. One of the attack vectors demonstrated was physical USB access (perhaps delivered by a disgruntled employee), with malware designed to manipulate the training, validation and testing data. This malware was undetectable by Windows 11 AV, given its unique heuristics. The poisoned backdoor attack was one of the most effective attacks against the MAS AI model. In one scenario, an attacker waits until an autonomous vessel is at a significantly vulnerable point, such as a busy port entrance or in a narrow water channel. Then, the attacker displays the hidden trigger physically on or near an object, which confuses the MAS into misidentifying or not identifying a particular marine object (such as an oil tanker); the MAS collides with the vessel, causing the port or channel to stop operations at significant disruption and cost. For perspective, the non-cyber related grounding of the Evergiven container ship in the Suez Canal caused disruption of approximately \$54 billion in trade loss \citep{lee2021suez}. Equally blue team scenarios for security using a backdoor patch also exist to defend against hostile drones. 

We then considered multiple evasion-based attacks against the MAS. Due to a lack of AAI defenses, we showed numerous attacks were also theoretically possible, such as FGSM for open-box and EA-based attacks for closed-box access. However, practical implementation of these attacks would be challenging for a USV. The generation of the attack would require changes to the input (camera frames), and so would require access to modify the model input channel. Furthermore, the generation of effective perturbations can take significant compute time, and with a refresh rate of 60 FPS, this attack would be challenging against a USV of this configuration. Patch-based attacks appeared more promising, with strong transferability between models. The patch-based attacks could be used in a similar way with the same effect as the backdoor poison attack; however, they do not require the initial stage of manipulating the model training data.

Inference and inversion-based attacks do not appear as relevant to this application of AI. The anticollision AI model is easy to recreate and not trained on any sensitive data. However, subject to changes in the future, this attack may become more relevant, so we briefly discuss this attack and its mitigations in this work. This highlighted the importance of AI red teaming to find vulnerabilities and implementing strong adversarial AI defenses, such as poison and anomaly detection, input sterilization and validation, adversarial training, and to utilize multimodal model input (multiple sensors) (see the seven secure principles of \cite{walter2023adversarial}); this makes the task of these attacks more challenging with a greater number of sensors to manipulate. Depending on how the sensors are integrated could determine the difficulty of a successful attack. For example, an attacker could cause a DoS to the sensors, resulting in an overreliance on a single sensor, which could be attacked depending on the integration and weighting method of the sensors.

\section{Conclusion}
\label{sec:Conclusion}
Previously, limited uptake in real-world AI technologies and a lack of understanding of the real-world AAI threat landscape has meant the threat of AAI was perceived as somewhat theoretical. Nowadays, the threat landscape is considered fast growing and evolving with the ubiquitous uptake of AI to enhance a diverse range of tasks, greater accessibility to AI systems and greater dependence on AI for mission-critical systems. 

In this work, we have seen how vulnerable a real-world MAS is to numerous adversarial attacks by proposing a red team framework, RED-AI, to evaluate the adversarial attack surface of a real-world MAS. RED-AI was used to identify numerous vulnerabilities to AI models with a range of lethality and discussed effects and mitigations, showing its high effectiveness when used to red team AI systems in high entropy, dynamic, real-world environments.  In future work, we will consider multimodal input AI systems analyzing adversarial attacks against radar, image, FLIR and AIS dependent MAS, developing more advanced threats against more advanced systems.

RED-AI provides operators with more than just a proactive response for securing by design of new AI technologies, but also a reactive response to securing existing AI technology today and in the future as threats evolve. Furthermore, using RED-AI to evaluate the security of AI (MAS) models not only helps identify AAI threats to the MAS but also can increase the system's robustness, which could benefit the technology in complex scenarios, adverse weather conditions and better detection through increasing greater model robustness. 

Ultimately, the proposed RED-AI framework aims to allow a better understanding of the real-world AI threat landscape, supporting one to develop better attacks and defenses against real-world AI. Although we have evaluated MAS, the same lessons can be applied and transferred to other non-MAS mission-critical AI systems. RED-AI provides cyber and AI practitioners with a tool to better protect their assets from the growing and looming adversarial threat.

\section*{Acknowledgements}

This work was supported by the Turing’s Defence and Security programme through a partnership with the UK government in accordance with the framework agreement between GCHQ and the Alan Turing Institute. The authors would like to thank the University of Plymouth for the use of their autonomous fleet. The authors would also like to extend their gratitude to David Bowman for his support throughout the deployment process.

\section*{Disclosure statement}

The authors report there are no competing interests to declare.

\section*{Data Availability statement}

Due to the sensitive nature (security) of the research, supporting data is not available.













\bibliographystyle{tfcad}
\bibliography{main}

\vspace{-0.0cm}
\section{Appendices}


\appendix

\section{RED-AI Checklist}
\label{append:checklist}
This section provides a checklist to support the RED-AI framework to evaluate AI systems. Not all these check marks and sections may be relevant, and this checklist should be tailored for each evaluation. 

\subsection{Define the Scope of Evaluation}
\label{App:DefineScope}
\begin{itemize}
    \item A questionnaire or meeting to answer some of the questions defined below and provide a mutual understanding of the evaluation.
    \item Discuss the evaluation's objectives. Are we evaluating the whole lifecycle or post-development AI?
    \item Understand and look to minimise any possible disruption to the AI's owner. If something does go wrong, who do we contact?
    \item Discuss limitations of the test or any assumptions made during the evaluation.
    \item Gain appropriate permissions? Agree on access (including physical) to the systems? And on what schedule?
    \item How will the tests be carried out? In the deployed environment? Against an API? With software and hardware access? With the use of digital twins or simulators? With a system's physical clones? And on what schedule?
    \item Rules of engagement should be defined. What is on and off limits to evaluation?
    \item The disclosure process of vulnerabilities should be properly settled with the AI owner before enacting the evaluation. 
    \item Confidentiality of the AI systems? NDAs? Are there any relevant security classifications and required clearances?
    \item How easy will it be to obtain access to MAS software when considering built-in-house or third-party software use?
    \item In some applications, explainability and ethics could be considered.
    \item Will the AI owner want to observe any stages of testing? 
    \item Start and end date of the evaluation?
\end{itemize}

\subsection{Information Gathering and Threat Modeling}
\label{App:Info}
\begin{itemize}
\item Consider both external publicly available data (OSINT) and internal data relating to the owner's AI. 

 \item Investigate marketing and corporate material? Such as academic literature, GitHub repositories, websites and social media.
 \item Understand relevant documentation policies. Does the organization track the use and details of utilized AI systems?
 \item Obtain the AI software. Set up an environment to simulate and evaluate the software. Obtain any relevant hardware to support the evaluation.
 \item Understand any third-party collaboration in AI development.
 \item Enumerate and gain access to development environments, containers, model zoos, development software, versions, models and architecture.
 \item Identify any higher-risk techniques used, such as transfer learning, the reuse of public data and continuous learning.
 \item If public datasets/models are reused, look for existing vulnerabilities and biases. Investigate known poisoning attacks against the models and datasets.
 \item Consider the use and type of datasets utilized by the AI model. Consider the confidentiality of the data used.
 \item Investigate data collection techniques and validation? 
 \item Investigate model and data decommissioning standards?
 \item Enumerate sensors and software utilized by the AI? Understand the technology and sensor's relationships. 
 \item Investigate how, by whom and when the model was trained. 
 \item Investigate the input and output interfaces for models. Are input sanitation techniques utilized? Does the model provide explainability? Does the model provide confidence predictions or exact values?
 \item Does the AI utilize any defenses? Does the AI utilize sensor redundancy? Does the AI utilize multiple sensors as AI model input?
 \item Perform a risk assessment of the AI. What are the policies and understanding of AAI risks within the organization? 
 \item One could also consider the physical security of the AI.
\end{itemize}

\subsection{Evaluation}
\label{App:Eval}
Using information obtained in the data gathering stage, one can now evaluate the model's lifecycle and/or evaluate the model post-deployment.

\subsubsection{Evaluation throughout the model's lifecycle}
\begin{itemize}
 \item Evaluate the security of the model throughout its lifecycle.
 \item Assess the quality of data collection practices. Such as encryption, storage, access and transfer.
  \item Investigate relevant CVEs for development software and environments.
  \item Assess model training practices.
  \item Assess the evaluation of the model's performance. Is the model's robustness and ability to defend against AAI attacks considered in the evaluation process?
  \item Assess the decommissioning of the model and data practices.
  \item Is there a `secure by design' approach to AI development?
\end{itemize}

\subsubsection{Deployment evaluation - inversion, evasion, poisoning and extraction attacks}
This stage is application-specific and depends on the access to the model, for example, open-box or closed-box attacks. Check threat intelligence sources for novel AAI attacks against AI models before performing the evaluation. Section~\ref{sec:TestCase} contains examples and further literature on attacks and defenses for inversion, evasion, poisoning and extraction attacks, which may support the evaluation.
\begin{itemize}

\item Determine which attacks are relevant for the AI access, model and application under evaluation.
\item One may consider conventional cyber security attacks, such as attempting to bypass security measures implemented within the AI system. The use of prompt injection. A DoS of sensors and communications.
\item Test the effects of attacks on different sensors, inputs and AI components.
\item Test and collect results from evasion attacks.
\item Test and collect results from inversion attacks.
\item Test and collect results from stealing attacks (inference and inversion).
\item Test and collect results from poisoning attacks.
\item Test and collect results from closed-box attacks.
\item Test and collect results from open-box attacks. One may also consider combining previous successful attacks, e.g., utilizing a proxy model from an earlier model extraction attack to create more accurate evasion attacks (open-box).
\item Test and collect results from any relevant novel AAI attacks. AAI Github repositories and existing threat intelligence methods may be one way to keep current with the latest AAI methods.
 \end{itemize}

\subsubsection{Scenarios of Attacks}
\label{App:Scenario}
This section should be used to test the ability of the technology to be robust against attacks whilst considering the impact and likelihood of attacks. One could use this information to consider scenarios and prioritize threats based on their likelihood and greatest gain (impact) to the attacker. 

\begin{itemize}
\item Assess the ability of the system to recognize spoofed sensors and respond to attacks or sensor failures.
\item Consider the monitoring/logging queries' usefulness in an attack.
\item Examine the resilience of sensors to attacks and failure. Are any sensor defenses implemented (or redundancy), and what is the effect during attacks? 
\item Consider the limitations of the employed defensive measures. 
\item Consider the misuse of the technology.
\item Consider the possibility and likelihood of attacks and threats, scenarios for which the threat could be at greatest risk or of greatest gain to the attacker.
\end{itemize}

\subsection{Reporting and Mitigation}
\label{App:Report}
Provide a report (and a meeting) of the actions conducted during the evaluation and the findings. Items the report could include:
\begin{itemize}
\item Disclosure of potential vulnerabilities.
\item Any relevant threat metrics for evaluation.
\item The limitations and assumptions of any defensive measures. 
\item Recommended mitigations.
\item Possible effects of attacks and threat scenarios.
\item Highlight any areas one could not fully evaluate.
\item Check against the NCSC's principles for the security of machine learning \citep{anonymous_2022} and recommend these NCSC principles where appropriate.
\end{itemize}

\subsection{Validation and Retesting}
\begin{itemize}
\label{App:Val}
\item Validation and then retesting, if appropriate.
\item Consider periodically retesting AI systems as the threat landscape evolves.
\end{itemize}

\end{document}